\documentstyle[11pt]{article}
\begin{document}
\pagestyle{plain}
\huge
\title{\bf Missing experiments in relativity and gravity}
\large
\author{Miroslav Pardy\\[7mm]
%Institute of Plasma Physics ASCR\\
%Prague Asterix Laser System, PALS\\
%Za Slovankou 3, 182 21 Prague 8, Czech Republic\\
%and\\
Department of Physical Electronics \\
and\\
Laboratory of Plasma physics\\[5mm]
Masaryk University \\
Kotl\'{a}\v{r}sk\'{a} 2, 611 37 Brno, Czech Republic\\
e-mail:pamir@physics.muni.cz}
\date{\today}
\maketitle
\vspace{5mm}

\begin{abstract}
The proposal is submitted for the measurement of the relativistic length contraction using the nanoscopic dumbbell moving in LHC or ILC. Some paradoxes concerning the length contraction including the relativity of simultaneity is  discussed. The dynamical state of rods and strings accelerated by the gravitational field and non-gravitational field is discussed. The definition of the length in the rotating system is not considered here.  The realization of the acceleration of the nanotube charged dumbbell in LHC, ILC and other accelerators is evidently prestige problem for the experts in the accelerator physics. 
\end{abstract}

\vspace{3mm}


\vspace{7mm}

\section{Introduction}

Experiments in physics involves measurement and observation. While  terrestrial physics enables to change the conditions of the measurement, we perform in  astronomy the observation only and we do not  change the experimental conditions.
Considering the length contraction, we can perform the measurement of the length of the moving rod, or the observation of the same, or observation of some moving object. However at this moment the observation of the length must be strictly distinguished from the measurement of the length as was shown for instance by Terrell (1959), Weisskopf (1960 ) and others (Dreissler, 2005). We believe that the same is  valid for the time dilation.
The measurement of the moving object gives the contraction by the factor $\sqrt{1 - v^{2}/c^{2}}$ in the direction of motion for a moving object in comparison to the length in the rest frame of the object. It would be incorrect to state that we
"see" the length contraction, or, that the length "appears" to be contracted by the factor   $\sqrt{1 - v^{2}/c^{2}}$. 
 As first pointed out by Lampa (1924) and later by
Penrose (1959),  Terrell (1959) and Weisskopf (1960) what one "sees" and how
an object "appears" are very different from what is given by
the Lorentz contraction. The reason is that various parts of
the object are different distances from the observer, and in
order for the light rays from the various parts to arrive at the
observer at the same time, they must have left the object at
different times. It follows from the special theory of relativity that the length contraction is the result of the measurement procedure 
and the time dilation is also the measurement procedure as was shown by Fok (1961) and author (Pardy, 1969).

The photograph of a relativistically
moving object with a camera using, instead of photons,
particles moving much faster than the velocity of light, eliminates the non-desired optical effects and the film would
show the object shortened by a factor of  $\sqrt{1 - v^{2}/c^{2}}$  in the direction
of motion. However such a camera is not physically possible,
and we can ask how to correct for the optical effects so that only
the relativistic effects will be observed on a photograph taken by an ordinary camera. 

In this paper we refer to new approach the measurement of the Lorentz contraction.
We use the synchrotron spectrum of the rigid two-body system in such a way, the we read the information on the Lorentz contraction 
from  this spectrum as the proof of the Lorentz contraction. 

We frequently read that the Lorentz contraction of moving object is only relative
contraction as the opposition of the Fitz-Gerald contraction of moving object 
with regard to the aether. So, it also means that the carbon 
nanotube moving in the quark machines (Fraser, 1997),  i.e. in Large Hadron Collider (LHC), or, International Linear
Collider (ILC)  is contracted with regard to LHC, or ILC. The nanotube is
conductive object of many carbon atoms. Every atom can be ionized and
it means every nanotube can be ionized. In other words, after
ionization, the nanotube is charged and can be accelerated. The charged
nanotube produce the \v Cerenkov radiation if moving in dielectric
medium, or synchrotron radiation when moving in the magnetic field. The information on the Lorentz
contraction is evidently involved in the spectrum of the \v Cerenkov
radiation, or, in the spectrum of the synchrotron radiation when the
nanotube moves in LHC, or in ILC. The orientation of the nanotube during the
motion is not theoretically elaborated and no experiment was performed
to investigate this problem. The spectral formulas
for the synchrotron radiation of the carbon nanotube 
are not very simple because the nanotube is composed of  the great
amount of elementary charges which corresponds to the sum  of 
many delta-functions for the charge density of this object.

However, the problem is possible to modify if we replace the nanotube by the
the two  nanotube segments connected by the silicon nanorod. At
present technologies this operation is the feasible problem. After ionization  of segments, every segment can be considered as the point charge and the carbon-silicon-carbon is the two-body charged system.  In such
a way we reduce the many-body problem to the two-body problem where the
system is the one-dimensional element with the charged ends. It means we can
apply the derived formulas for the two-body problem to our C-S-C
system, where C is a carbon and S is a silicon. The classical analogue of
this system is the macroscopic dumbbell created from the two golf balls fixed together by the thin metal rod. 
We suppose that the  carbon segments are multiwall nanotubes and the silicon
nanorod connecting the ends is dielectric. The left end and right and of the silicon nanorod is fixed 
inside the cavity of the carbon nanotube. Such system is prepared
to verify the Lorentz contraction by LHC, or  by ILC. 
Use of the carbon is not necessary. We can us metal elements of the periodic table, such as Al, Fe, Cu, Sn, Au, Ag, and so on. 
If we use the silicon nonorod, we can fixed the atoms of the metals to the end of the nanorod by the method called sputtering which is the standard method of the nanorod physics. After ionization of the metal ends, the metal-silicon-metal nanorod behaves as the rigid two-body system and can be used at the experiments in LHC, or, ILC.    

\section{The length contraction paradox}

The well known is the Ehrenfest length contraction paradox (Sama, 1972). Let $R'$ be the radius of a cylinder rotating with regard to some rest system. Then $R'$ must satisfy two conditions that are in the contradiction.
(a) The circumference of the cylinder must show contraction relative to the stator. Or, 
$2\pi R' < 2\pi R$, since each element of a rotating cylinder is contracted.
(b) The element of radius is not contracted because there is zero relative velocity of radius with regard to the stator. 
So, we have two contradictory statements: $R' < R, R' = R$. The solution of this contradiction is elementary. Lorentz contraction is the result of the special procedure measurement. And, it is not possible to establish the procedure measurement for whole cylinder because the procedure measurement is based on the Lorentz transformation which is valid only for linear motion and  not for rotation. So, we have no contradiction but only a paradox.

If the system of axioms in STR is not contradictory than every logical consequence of this system of axioms is right. If the system of axioms of STR is contradictory, than we can get many absurd statements. The investigation of relativity axioms as  
the non-contradictory  system was to our knowledge not published. We believe that the relativistic system of axioms is not contradictory.

\section{The experimental verification of the length contraction}
  
The contraction of the length in the special theory relativity was
never experimentally verified. There is only the experiment and theory 
that the moving objects are turned with regard
to the observer (Terrel, 1959; Weisskopf, 1960). Now, living in the carbon century and 
nanotechnology century, we are able to
prepare the experiment with the carbon multiwall nanotubes, or metal-silicon nanotubes   which
enables to verify the length contraction when they are moving in a
such accelerator and collider as LHC, or ILC.

The method of verification is modification of the author older idea 
that the length contraction can be verified by 
the two-body system moving in the dielectric medium, or in the synchrotron.
The distance of two free charges moving in the dielectric medium is relativistically 
contracted and can be
obtained from  the \v Cerenkov spectrum of such two-body system. The basic formulas of such system was derived by author (Pardy, 1997; 2000)
and discussed by Cavalleri and Tonni (Cavalleri et al., 2000). 

Instead of such two-body system we consider the charged multi-wall
carbon dumbbell moving in LHC, or, in ILC, or, in dielectric medium and emitting the synchrotron radiation, or the \v Cerenkov radiation. The
carbon dumbbell is charged in such a way that the charges (approximately point like) 
are at the end
of the dumbbell. The realization of the dumbbell is a feasible problem for the carbon experts. 

If the carbon-silicon (metal-silicon) bi-pole (not dipole) is moving with the velocity $v$ in
the LHC and the rest
length is $l_{0}$, then the synchrotron spectrum of the harmonic of this bi-pole is
given by the formula (Pardy, 2000)

$$P_{bi-pole}(\omega, v, l) =  \cos^{2}\left(\frac {\omega l}{2v}\right)P_{mono-pole},\eqno(1)$$
where

$$l = l_{0}\sqrt{ 1 - v^{2}/c^{2}}\eqno(2)$$ 
and  $l_{0}$ is the rest  distance between charges. So we see that the resulting intensity of radiation is not the sum of the intensities of the individual charges, but it is result of the synergic process, which is not the classical interference.
 At the same time the mathematical form of the formula $P_{bi-pole}$ cannot be derived 
as the result of the interference of the different coherent source of the synchrotron light. The formula  $P_{bi-pole}$ is the consequence of the synergic process of the formation of radiation by the two charges at the ends of the dumbbell. The fact that the synchrotron spectrum of the dumbbell is of the modulated form of the one-charge spectrum is the new interesting physical synchrotron radiation reality which will be probably used in solid state physics, chemistry, biology and other scientific disciplines.

Author used the method of deceleration (Pardy, 1997) of the two-body system, in order the get the rest length, 
however his method is very sophisticated.
 In case of the carbon-silicon (metal-silicon) dumbbell, the deceleration can be realized by means of the bremsstrahlung loss of the charged   dumbbell moving in the magnetic field of LHC. Then, the relativistic velocity of a dumbbell is the variable experimental quantity.

The more simple formula was derived in case that the two-body system
of charges was moving in the dielectric medium with the velocity
sufficient to produce the \v Cerenkov radiation. The formula is as
follows:

$$P(\omega,v, l) = \cos^{2}\left(\frac {l\omega}{2v}\right)
\frac{e^2}{\pi}\*\frac{\mu\omega}{c^{2}}
v\left[1 - \frac{1}{n^2\beta^{2}}\right]; \quad n\beta>1 \eqno(3)$$
and

$$P(\omega,t) = 0;\quad n\beta < 1 .\eqno(4)$$

The zero point of function $P(\omega,t)$ are as follows:

$$\omega_{0} = 0;\quad \frac {\omega_{n} l}{2v} = \frac {(2n-1)}{2}\pi;
\quad  n = 1, 2, 3, \dots  .\eqno(5)$$

From the last equation follows

$$l = \frac {(m-n)2\pi v}{\left(\omega_{m} - \omega_{n}\right)} =
l_{0}\sqrt{1-\frac {v^{2}}{c^{2}}},\eqno(6)$$
or,

$$l_{0} = \frac {2\pi v}{\sqrt{1-\frac {v^{2}}{c^{2}}}}
\frac {(m-n)}{\left(\omega_{m} - \omega_{n}\right)}.\eqno(7)$$

If we know the $n$-th and $m$-th zero points with the corresponding $\omega$-s
and velocity of the charges we can exactly determine their rest distance.
Then, the rest distance determined by the formula (7) can be compared with
the rest distance of the charges obtained by direct measurement and in such
a way we can verify the Lorentz contraction. The velocity is in experiment variable quantity.

We get for different velocities the different bremsstrahlung spectra
and we can these spectral formulas compare in order to show that the radiation
depends on the contracted length and not only on the rest length.
The measurement of the bremsstrahlung spectra  enables to determine
and verify the direct Lorentz contraction of length.
We do not write down the spectral formula generated by moving dumbbells in ILC.

In case of two charges the problem was how to slow down the
velocity. It was discusses by author (Pardy, 1997). The sophisticated approach was evident. The slowing down by the bremsstrahlung of the two-body
system was also problematic. The similar problem is the Bell fictitious experiment with the two identical rockets with 
the identical acceleration where the initial distance between them is $l_{0}$. The problem is still under discussion. 
Modification of this problem is a such that  the motion of a rod is caused by the same acceleration at the ends. What is the dynamical state 
of the rod in $S$ and $S'$? This problem of equations of mathematical physics was never solved.  

However, we can use the carbon bi-pole dumbbell, where the rest length of the
dumbbell is constant instead of the two free charges.
The slowing down of the dumbbell in the LHC is performed by the
bremsstrahlung process. According to Schwinger (1945), the energy loss (the bremsstrahlung loss) is 20 eV per 
revolution of an electron with energy $10^{8}$ eV and radius 0.5 m.
In such a way the problem of the Lorentz length contraction is solved. 

The orientation of he carbon bi-pole is tangential with regard to its trajectory 
as  an analogue of the  mechanical dumbbell connected by fibers to the center of rotation 
and rotating around it.

In case  we use the linear accelerator
then, the carbon segments must be of the different masses in order to get the stability of the dumbbell on the trajectory.
We know from the history of spacemanship, that the Soviet module rotated during the deceleration in the Earth atmosphere and therefore it was not possible to stop this module rotation by the parachute.

\section{Accelerated rod and string and gravity}

The experimental investigation of the accelerated rod or length was not to our knowledge performed. Nikoli\v c (1999) discussed the dependence of length on acceleration. Author (Pardy, 2004; 2005) proposed so called nonlinear Lorentz transformation which is based on the ad hoc principle valid for the acceleration systems. The definition of the length in the rotating system is not considered here. Only bound motion of bodies in rotating system was considered by author (Pardy, 2007a).  Nikoli\v c and author do not consider the internal structure of the rod, or string. The meaning of the present discussion is to consider the accelerated rod as the physical system with the internal structure, which enables to say that the gravitational field differs from the the non-gravitational field.

1. The dynamics of the rod, or string accelerated by the gravitational field differs from the dynamics of ones accelerated by the non-gravitational field. If the rod, or string is hanged in the gravitational field, then their length is greater than the length in
zero gravitational field. Then, if the end is free, the  free fall occurs. The internal motion of these systems follows from the equations of mathematical physics. 
 If the rod, or string is accelerated (from left to right) by force fixed to the left end of them, then the length is contracted. The internal motion of the rod, or string is determined by the equations of mathematical physics. However the internal dynamic of the systems accelerated by the gravitational field differs from the internal dynamics of the systems accelerated by the non-gravitational field.   
2. It follows from 1. that we are able to distinguish the gravitational field from the non-gravitational field by experiment.
3. It follows from 1. and 2. that the principle of equivalence is not valid for accelerated rods, or strings.
4. It follows from 1., 2., 3. that the validity of the principle of equivalence is not absolute. 

The goal of the Stanford experiment is to confirm that bodies with the different mass fall by the same acceleration (Galileo proved it by his experiment performed in Pisa). The Stanford experimenters have at this time opportunity to verify that the dynamical states of rods and strings (we mean real strings) differ when accelerated by gravitational field from them accelerated by non-gravitational field. The analogical  experiment can be performed using the linear chain falling in the gravitational field of the Earth. Co-moving CCD camera can register the internal motion of the chain. 
At this time the experiments with accelerated rods and strings are the missing experiments of gravity physics. The Einstein gravity must  be in such a way completed by the informations following from the new experiments with rods, strings, or linear chains.    

\section{Discussion}
 
The acceleration of the dumbbells by LHC and ILC is the probably the future goal of these quark machines. 
Such goals are not still involved in the monographs on accelerators (Fraser, 1997).
 
The  first accelerator was built for the realization of the reaction $ electron + positron \to anything$. The LHC is built for the reaction $ proton + proton \to anything$. The possible generalization of the last process is the reaction $ion + ion \to anything$.
And we submit the proposal 

$$dumbbell  + dumbbell  \to anything. \eqno(8)$$

We believe that such processes described by the last equation are feasible and that they have the deep physical meaning for the particle physics in order to get new information about matter. There are many particle accelerators around the world accelerating electrons, positrons, protons and so on ($http://www{-}elsa.physik.uni{-}bonn.de/accelerator_{-}li$). We hope that some of them are suitable 
to involve the acceleration of dumbbell with the goal to verify  experimentally the Lorentz contraction, 100 years after the appearance of the special theory of relativity.        

The synchrotron radiation evidently influences the motion of
the electron in accelerators. The corresponding equation which describes
such motion is so called Lorentz-Dirac equation, which
differs from the the so called Lorentz equation only by the additional
term which describes the radiative reaction (Landau et al., 1988):
 
We elaborated the dumbbell theory of the length measurement by LHC and ILC  because the theory where two electrons were  used  was very sophisticated (Pardy, 1997; Cavalleri et al., 2000). On the other hand, there is still the missing derivation of the dumbbell synchrotron spectrum in the framework
of the S-matrix theory of the QED and with regard to the knowledge of the Volkov solution of the Dirac equation (Pardy, 2007b). The solution of this problem involves the derivation of the Volkov solution for the Bethe-Salpeter equation for the two-body problem where the two-body system is the dumbbell. We hope that this missing problem will be solved in the near future.

The problem of the deceleration of the electron-electron system was the problem of the simultaneous deceleration. However,  we  proved in the different article that the simultaneity as such leads to the paradoxes (Pardy, 2007c). The notion  of the simultaneity
in case of the deceleration of the dumbbell plays here no problem because the dumbbell is the rigid system in the rest and in the tangential orientation to trajectory of motion in accelerators. 
  
Let us show that the Einstein realization of simultaneity (Einstein, 1905; 1919) leads
to the paradox if we consider the dynamics of the elastic string. 

So, let the string in the system $S$ with the equilibrium length $l_{o}$
is elongated to the length $l$, $l  > l_{o}$, and it is fixed et the
ends. At time $t= 0$ the ends of he string are released. The
string motion is described by the wave equation 

$$\frac{1}{a^{2}}\frac{\partial^{2} u(x,t)}{\partial t^{2}} - 
\frac{\partial^{2} u(x,t)}{\partial x^{2}} = 0 \eqno(9)$$ 
including the initial conditions

$$u(x, 0) = Ax; \quad \partial u(x,t)/\partial t|_{(t=0)} = 0; u(x, 0) \eqno(10)$$
and the boundary conditions expressing the fact that the ends of the
string are free from time $t > 0$. Here $a \neq c$, $c$ being the 
velocity of light.

Without mathematics we know that the center of mass of string is at
the rest. This result can be immediately confirmed using the rubber
string with the fixation of ends by fingers. In other words, by the
string-finger experiment. The experiment with the rubber string is
very simple and can be performed easily.

The situation in the system $S'$ moving with the velocity $v$ with
regard to the system $S$ is different because the releasing of the
ends of the rubber string is not simultaneous according the Einstein
definition of simultaneity. 
It means that the motion of the string is a such that
the center of mass changes its velocity. We also can verify the change
of velocity of the center of mass by the rubber string experiment
with the non-instantaneous releasing of the ends of the string.

So, we have a paradox. The center of mass of he string in system $S$ does
not change its state of motion, while in the system $S'$ it does. This
paradox  is not involved 
in the collection of paradoxes of  relativity (Goldblatt, 1972;
Terletzkii, 1966) and in the relativistic  paradoxes in American Journal of
Physics. To our knowledge, this paradox is not involved in any
monograph of the string theory. The resolution of this paradox is described in the very simple way in the author e-print (Pardy, 2007c).

\newpage
%\vspace{15mm} 
\noindent
{\bf References}.

\vspace{10mm} 

\noindent
Cavalleri, G. and Tonni, E., (2000). Comment on \v Cerenkov effect 
and the Lorentz contraction, Phys. Rev. {\bf A 61}, 026101. \\[2mm]
Einstein, A., (1905). Zur Elektrodynamik bewegter K{\"o}rper, Annalen der
Physik, {\bf 17}. See also {\it The Principle of Relativity}, 
published in 1923 by Methuen and Company, Ltd. of London. Most of the 
papers in that collection are English translations by W. Perrett and 
G.B. Jeffery from the German {\it Das Relatitivit{\"a}tsprinzip}, 
4th ed., published in 1922 by Teubner.  \\[2mm]
Einstein, A., (1919). {\it \"Uber die spezielle und die allgemeine
relativit{\"a}ts Theorie}, Vierte Auflage, Vieweg {\&} Sohn, 
Braunschweig.; chapter 9.\\[2mm]
Fraser, G., (1997). {\it The quark machines}, (Institute of physics
Publishing, Bristol and Philatelphia). \\[2mm] 
Dreissler, R. J., (2005). The appearance, apparent speed, and removal of optical effects
for relativistically moving objects, Am. J. Phys. {\bf 73} (7), July. \\[2mm]
Fok, V., (1961). {\it Theory of space, time and gravity}, GIFML, Moscow. \\[2mm]
Goldenblat, I. I., (1972). {\it The time paradoxes in the special
theory of relativity}, (NAUKA, Moscow), (in Russian). \\[2mm]
Landau, L. D. and Lifshitz, E. M., (1988). {\it The classical theory of fields}, 7th ed.,
(Moscow, Nauka), (in Russian).\\[2mm]
Lampa, A., (1924). Wie erscheint nach der Relativit{\"a}tstheorie ein bewegter Stab
einem ruhenden Beobachter? (How does a moving rod appear for an
observer at rest according to the theory of relativity?), Z. Phys. {\bf 27}, 138--148. \\[2mm]
Nikoli\v c, H., (1999). Relativistic contraction of an accelerated rod,  Am. J. Phys.
{\bf  67}, 11, November. \\[2mm]
Pardy, M., (1969). A remark on the clock paradox, Phys. Lett. {\bf 28 A}, No. 11, 766. \\[2mm]
Pardy, M., (1997). \v Cerenkov effect and the Lorentz contraction, 
Phys. Rev. {\bf A 55}, No. 3, 1647. \\[2mm]
Pardy, M., (2000). Synchrotron production of photons by the two-body system,
Int. Journal of Theoretical Physics {\bf 39}, No. 4, 1109; hep-ph/0008257. \\[2mm]
Pardy, M., (2004). The space-time transformation and maximal acceleration,
Spacetime {\&} Substance Journal, {\bf 1}(21), pp. 17-22. \\[2mm]
Pardy, M., (2005). Creativity leading to discoveries in particle physics and relativity; ibid. physics/0509184.\\[2mm]
Pardy, M., (2007a). Bound motion of bodies and particles in the rotating systems,
International Journal of Theoretical Physics {\bf 46}, No. 4, April 2007; 
Revised and modified version of astro-ph/0601365. \\[2mm]
Pardy, M., (2007b). The synchrotron radiation from the Volkov solution of the Dirac equation,
hep-ph/0703102. \\[2mm]
Pardy, M., (2007c). New paradox in the special theory of relativity 
generated by the string dynamics, physics/0705024. \\[2mm]
Penrose, R., (1959). The apparent shape of a relativistically moving sphere,
Proc. Cambridge Philos. Soc. {\bf 55}, 137–139. \\[2mm]
Sama, N., (1972). On the Ehrenfest paradox, Am. J. Phys. {\bf 40}, 415. \\[2mm]
Schwinger, J., (1945). On radiation by electrons in betatron, LBNL-39088, CBP, Note-179, UC-414.\\[2mm]
Terletzkii, Ya. P., (1966). {\it The paradoxes of the special theory
of relativity}, (NAUKA, Moscow), (in Russian).\\[2mm]
Terrell, J., (1959). The invisibility of the Lorentz contraction, Phys. Rev. {\bf 116}, 1041. \\[2mm]
Weisskopf, V. F., (1960). The visual appearance of rapidly moving objects,
Physics Today {\bf 13}, 24. \\[2mm]
\end{document}